\documentstyle[preprint,aps,pre]{revtex} 
\begin{document}                          
\draft                                    

\draft
\title{A Simple Model for Anisotropic Step Growth}
\author{J. Heinonen$^{1}$, I. Bukharev$^{2}$, T. Ala-Nissila$^{1-3,*}$, 
and J. M. Kosterlitz$^{2}$}
\address{
$^{1}$Helsinki Institute of Physics, P.O. Box
9, FIN--00014 University of Helsinki, Helsinki, Finland
\\
$^{2}$Brown University, Department of Physics, Box 1843, Providence R.I.
02912--1843
\\
$^{3}$Laboratory of Physics, P.O. Box 1100, 
Helsinki University of Technology, FIN--02150 HUT,
Espoo, Finland
}

\date{February 11, 1998}

\maketitle

\begin{abstract}

We consider a simple model for the growth of isolated steps on a
vicinal crystal surface. It incorporates diffusion
and drift of adatoms on the terrace, and strong step and kink edge barriers.
Using a combination of analytic methods and Monte Carlo
simulations, we study the morphology of growing steps in
detail. In particular, under typical Molecular Beam
Epitaxy conditions the step morphology is linearly unstable
in the model and develops fingers separated by deep cracks.
The vertical roughness of the step grows linearly in time,
while horizontally the fingers coarsen 
proportional to $t^{0.33}$. We develop scaling arguments to
study the saturation of the ledge morphology
for a finite width and length of the
terrace. 

\end{abstract}

\bigskip
\pacs{PACS numbers: 68.55.Bd, 68.35.Fx, 82.20.Wt}

\section{Introduction}

Atomistically controlled growth of metal and semiconductor 
crystal surfaces constitutes an important field of research
both from technological \cite{Eag95}
and fundamental theoretical \cite{Bar95}
points of view. Among all the different
growth techniques, Molecular Beam Epitaxy (MBE) has
a special status since it can be very efficiently used to
produce growth in well defined layer-by-layer
growth mode. 
Experiments using the reflection high-energy electron diffraction
technique \cite{Lar88} indicate two main mechanisms of growth in
such cases: 
layer growth by nucleation and spreading of 2D islands
on a nominally flat substrate, 
and step-flow growth of a
substrate with steps. In the latter case, it is crucial
to be in the regime where the flux of adatoms is small
enough, and their diffusion fast enough to avoid island nucleation
on terraces. Such a window of the relevant 
physical parameters may be found experimentally for many
materials \cite{Eag95}. 

An important practical realization of the step-flow situation 
is epitaxial growth on a {\it vicinal} surface that
is cut in a direction slightly off from a high-symmetry
one. Such surfaces often 
consist of broad terraces of size separated by monoatomic
steps at distance $\ell$ from each other. 
By changing the miscut angle, the
density of the steps and thus $\ell$
may be controlled.  
The physics of MBE growth on such surfaces can be
in the simplest terms described by the following 
schematic model (see Fig. 1).
There is a flux $F$ of adatoms that impinges
upon the terraces. Particles on terraces then diffuse around
with an associated diffusion constant $D$,
and may be desorbed after a time $\tau$.
Upon approaching step edges, particles can either
cross them from above or below, be reflected from
them, or be incorporated into the growing ledge.
Attachment is usually characterized by
Arrhenius type rate constants $k_+$ and $k_-$ which refer to
the average rates of particles arriving at the ledge from
below or above, respectively.

This simplified picture of step flow growth was first
introduced by Burton {\it et al.} \cite{BCF}.
More recently, attention has been drawn to the
fact that in many real systems, $k_+$ and $k_-$
need not be equal \cite{Ghe88}
because of the existence of {\it step edge barriers}
\cite{Sch69}.
These barriers may often be present at step edges due
to reduced coordination of atoms. Recent theoretical
work shows that the step barriers play an important role
in controlling growth under MBE situations 
\cite{Kru93a,Sie94}.
In particular,
if these barriers are high adatoms cannot cross steps,
and the particle current will be in the direction of ascending
steps. For vicinal surfaces, this stabilizes the step flow
growth mode when nucleation on terraces is neglected.
If the average distance between nucleation centers is $\ell_N$,
step flow growth requires that $\ell/\ell_N \ll 1$. 
  
Most of the recent 
work dealing with step growth has concentrated on
the global properties and kinetic
roughening of growing surfaces with steps
under MBE conditions \cite{Eag95,Bar95,Growth}. 
However, attention has also been paid on the 
properties of individual steps under growth
\cite{steps,Bal90,Sal93,Pie96,Pie97}. It is a
well known property of ideal, isolated 1D steps that they are
thermally rough above zero temperature due to kinks.
Using linear stability
analysis, Bales and Zangwill \cite{Bal90} have shown
that in a system with unequal attachment rates $k_+ \not=
k_-$, a straight terrace ledge can become unstable when
$k_+ > k_-$. This kind of
growth-driven instability is particularly interesting 
since it may lead to the appearance of ``wavy''
patterns of the ledges.
More recently, Salditt and Spohn \cite{Sal93}
have argued that in addition to the instability, there is
a regime for isolated steps (in the case of
strong step edge barriers)
where the Kardar-Parisi-Zhang (KPZ) \cite{KPZ} theory
of kinetic roughening is valid. In this regime, the 
width of the ledge eventually roughens as $t^{1/3}$ in
analogy to many 1D surface deposition models \cite{Bar95}.

In this work, we study the nature of ledge or step edge 
morphologies, and the question of their roughening
behavior in a simple but nontrivial
model of isolated steps. This model is in part 
motivated by the energetics of adatoms
on Si(001) surfaces with widely separated steps. 
In the model, we assume
infinitely strong step edge barriers, and
biased diffusion both on the terrace and at the ledge.
As expected \cite{Bal90}, the ledge always becomes
morphologically linearly unstable due to the dominance of
the one-sided diffusion field.
Through a combination of analytic arguments and
computer simulations we show that
the ledge develops finger-like structures and its
roughness grows linearly in time, in contrast to the
KPZ type of roughening predicted by Salditt and
Spohn \cite{Sal93} in the stable regime. 
In addition, we study the
lateral coarsening of these fingers and show
that it follows a $t^{0.33}$ behavior.
We develop scaling arguments to study the influence of
the finite width and
finite length of the terrace on the growth.
Finally, we discuss the relevance of these results
with respect to experiments on
ledge roughening under MBE growth \cite{Zan93}.

\section{Anisotropic Step Growth Model}

\subsection{Definition of the Model}

The model is defined on a two dimensional square 
lattice where there is a single growing step. 
The average direction of the ledge
is along the $x$ axis where the
width is $L_x$ with periodic boundary conditions.
Initially at time $t=0$
the step at $y=0$
is completely straight with no thermal fluctuations
present. Growth of the ledge is initiated by depositing
a single particle randomly on an empty, randomly
chosen terrace site in front
of the ledge at $y > 0$.
After this, the particle performs random walk
and drifts towards the ledge by jumping
$\ell_d$ lattice sites in the $-y$ direction
at every random walk step on the average.
This means that during each step, the particle moves in the
$-y$ direction with a probability
$(1/4+\ell_d)/(1+\ell_d)$, while 
for the other three directions the probability is
$(1/4)/(1+\ell_d)$.

The ledge acts as an absorbing boundary
to the particle with the following rules (see Fig. 2): 
(i) if the particle arrives at the ``top''
(a section along the 
$x$ direction of the step), it is incorporated into it and becomes
immobile; (ii) if the
particle arrives at the ``side'' (a section
along the $y$ direction of the step),
it will instantaneously slide down along the ledge
to the $-y$ direction until it reaches the corner site where
it is permanently incorporated into the step. 
These rules guarantee that the set of
step heights $\{ h(x,t) \}$ 
as measured from $y=0$ obey the solid-on-solid restriction,
and the step forms a compact structure. 

After the particle has been
incorporated into the step, a new particle is deposited and the
process is repeated. Time in the model is measured in terms of
the average height of the growing step edge. We note that the size
of the terrace in the $y$ direction is not fixed, but is
chosen in such a way
that the distance from the highest point of the step
$H(t) \equiv \max{\{ h(x,t) \} }$ is kept at a constant value.
The corresponding boundary above $y=H+L_y+1$ is
completely reflecting and remains straight.
This means that a particle at $y=H+L_y$ that takes a step in the $y$ direction,
is immediately reflected back.

An important feature of the growth model is the
deposition of adatoms on the lower terrace only.
This is tantamount to assuming that the step barriers
are infinitely high with $k_{-}=0$ so that adatoms are reflected
from a downward step leading to an average particle current
in the $-y$ direction towards the up steps. The drift term $\ell_{d}$
is defined only in an average macroscopic sense and
will depend on the deposition flux and the concentration of adatoms
on the terrace in front of the step. Also, since we assume that
there is no desorption of adatoms ($\tau=\infty$), $\ell_{d}$ also
depends on the velocity of the step which in turn depends
on the terrace length $L_{y}$. Thus $\ell_{d}$ is, in
principle, determined self-consistently by the other parameters of
the model but we regard it as an independent parameter which may
be varied externally \cite{drift}. 

Finally, we would like to mention that
the growth rules of the model are in part motivated by
adatom dynamics on Si(001) surfaces with widely
separated steps \cite{steps2,Zha95,steps3,Gri89} 
under typical MBE conditions. Namely, on Si(001) diffusion
is spatially anisotropic both on the terrace
\cite{steps1,steps2,Zha95,steps3}
and at the step edges \cite{steps2,Zha95}.
However, at least for the case of single-height
steps on Si(001), microscopic
calculations \cite{steps2,Zha95} and experiments \cite{Zan93,steps3}
indicate that there is no significant step edge barrier.
Thus, we make no attempt to realistically model
the complicated adatom dynamics in this system, since
the main motivation here is to study the generic
features of the unstable regime for isolated steps.

\subsection{Simulation Algorithm for the Model}

A straightforward Monte Carlo simulation of the growth model proposed
here is in principle possible, but very difficult for large values of
$L_y$ and small drifts. This is because particles landing
on the terrace may wander arbitrarily far from the step edge,
and thus the time for a particle to become incorporated into
a growing step may become very large.
This problem can be solved by considering
the properties of 2D random walkers on a finite or semi-infinite
plane. For such cases, it is possible to calculate analytically
the spatial and temporal probability distributions for the walkers.
The idea then is that
for particles that initially land on the terrace
with $y>H+1$ (which is always the case if $L_y=\infty$), 
the simulation can be started by releasing them from
an imaginary line that runs along the $x$ direction just one lattice
site above the highest step, {\it i.e.} at $y=H+1$ (see Fig. 2).
If the particle crosses the line again in the $+y$ direction while
performing random walk,
it is immediately returned to it with a new $x$ coordinate chosen
from the appropriate spatial distribution which will be derived
below. In the Appendix we also
calculate the mean arrival (first passage) time of a walker 
and indeed show that this time becomes very large for small
values of $\ell_d$ and large $L_y$. 

More specifically, to implement the simulation algorithm
described above, we need to calculate
the spatial probability distribution function 
$P_{Lx,Ly}(x)$ which is used to obtain the new position
for a walker that crosses the line $y=H+1$ at any point. 
In other words, a walker
crossing the line being at $(x_0,H+2)$ with
any $x_0$, 
is put back to the new
site $(x-x_0,H+1)$ with the probability $P_{Ly}(x)$
where we assume for simplicity that $L_x=\infty$
(see Fig. 2).
For a discrete walker, this function satisfies the recursion
relation

\begin{eqnarray}
\label{Plyx}
P_{Ly}(x) & = & {1\over 2a}[b\delta_{x,0}+P_{Ly}(x-1)+P_{Ly}(x+1)+
\nonumber \\ 
& & \sum_{y=-\infty}^{\infty}P_{Ly}(x-y)P_{Ly-1}(y)],
\end{eqnarray}

where $a=2+2\ell_d$ and $b=1+4\ell_d$. 
Using the standard Fourier transformation

\begin{equation}
\label{Plyk}
\widetilde P_{Ly}(k)=\sum_{x=-\infty}^{\infty} e^{ikx}P_{Ly}(x),
\end{equation}

we obtain

\begin{eqnarray}
\label{Plyk2}
\widetilde P_{Ly}(k)& = &{1\over 2a}[b+e^{ik}\widetilde P_{Ly}(k)+
e^{-ik}\widetilde P_{Ly}(k)+
\nonumber \\ 
& & \widetilde P_{Ly-1}(k)\widetilde P_{Ly}(k)].
\end{eqnarray}

This gives

\begin{equation}
\label{contfrac}
\widetilde P_{Ly}(k)={b \over 2a - 2 \cos k - \widetilde P_{Ly-1}(k)},
\end{equation}

which must be solved with the initial condition $\widetilde P_0(k)=1$
for any $k$. For $L_y < \infty$, the continued fraction expansion
of Eq. (\ref{contfrac}) must be solved numerically in general. 
Even in the zero drift case the expansion converges rapidly, as
discussed in the Appendix. In the special case of an infinitely
long terrace $L_y=\infty$, 
$\widetilde P_{Ly}(k)=\widetilde P_{Ly-1}(k)$, 
and Eq. (\ref{contfrac}) gives

\begin{equation}
\label{Pinfty}
\widetilde P_{\infty}(k)=a-\cos k -\sqrt{(a-\cos k)^2-b}.
\end{equation}

In Fig. 3 we show the behavior of $P_{L_y}(x)$ for
various values of $L_y$ and $\ell_d$. In the continuum limit, 
the tail of this
function goes as $x^{-2}$ for the case of zero drift.

In practice, we also need the propagator for a periodic system with a finite
width $L_x$. This is most easily obtained in the Fourier space
by

\begin{eqnarray}
\label{Plxly}
& & P_{Lx,Ly}(x)=
\sum_{r=-\infty}^{\infty}
P_{Ly}(x+rL_x)
\nonumber \\
& &=\sum_{r=-\infty}^{\infty}
{\displaystyle 1\over{2\pi}}
\int_{0}^{2\pi}dk e^{-ik(x+rL_x)}\widetilde P_{Ly}(k)
\nonumber \\
& &=\int_{0}^{2\pi} dk e^{-ikx} \widetilde P_{Ly}(k)
{\displaystyle 1\over{2\pi}} \sum_{r=-\infty}^{\infty} e^{-ikrL_x}
\nonumber \\
& &=\int_{0}^{2\pi} dk e^{-ikx} \widetilde P_{Ly}(k)
{1\over{L_x}}\sum_{n=-\infty}^{\infty} \delta(k-{{2\pi n}\over{L_x}})
\nonumber \\
& & ={1\over{L_x}}\sum_{n=0}^{Lx-1}  e^{-i{{2\pi nk}\over{L_x}}}
\widetilde P_{Ly}({{2\pi n}\over{L_x}}).
\end{eqnarray}

Numerically, Eq. (\ref{Plxly}) 
is easy to implement using the Fast Fourier Transform
algorithm.

\subsection{Continuum Limit of the Model}

It is relatively straightforward to write down a continuum
description
by using the diffusion equation (an electrostatic analogy \cite{Witt81}
can also be employed). 
The probability density of a random walker
$u(\vec r,t)$ obeys the biased diffusion equation with a source
term $\rho(\vec r,t)$:

\begin{equation}
\label{continuum}
- \nabla \cdot {\bf D} \cdot \nabla u(\vec r,t)
+ \vec v \cdot \nabla u(\vec r,t) = \rho(\vec r,t).
\end{equation}

Using the distribution of the biased random walk, we can derive
expressions for the drift term $\vec v=(0,v_d)$ and
the diagonal elements of the diffusion tensor 
${\bf D}=D_{\mu \nu}$ ($\mu,\nu=x,y)$ to be
$v_d=-\ell_d/(1+\ell_d)$, and
$D_{xx}=1/(4+4\ell_d)$ and
$D_{yy}=1/(4+4\ell_d) + 8\ell_d/(4+4\ell_d)^2$.
We note that in the model, diffusion is always only slightly
anisotropic for $\ell_d>0$, and $D_{xx}/D_{yy}=1/3$ for
$\ell_d \rightarrow \infty$ \cite{irrelevant}.
The source term $\rho$ in Eq. (\ref{continuum})
is constant over the whole terrace.
The boundary conditions are that for the step edge $u=0$,
and for the reflecting boundary $\partial u / \partial y = 0$.
Also, the arrival probability of a random walker at the step edge
is proportional to the normal derivative of the probability field $u$.
With zero drift ($\ell_d=0$), Eq. (\ref{continuum}) reduces
to the Poisson equation obeyed by many growth models
(see {\it e.g.} Refs. \cite{Witt81,Kru93b,Kru97}).
The present sticking rules guarantee that the
growing step forms a compact structure, in contrast to the
typical Diffusion Limited Aggregation models \cite{Witt81}.
It is also evident from the 
stability analysis of Salditt and Spohn \cite{Sal93}
(see also Ref. \cite{Bal90}) that 
the one-sided diffusion field is
highly destabilizing, and the step edge
morphology is always controlled by the instability
rather than described by the nonlinear KPZ equation 
\cite{Sal93}.

\section{Numerical Results}

\subsection{Ledge Roughness}

We have performed extensive Monte Carlo simulations of the model 
with the algorithm described in Sec. II. 
In this work, we consider the case of finite $\ell_d$ only
\cite{zerodrift}.
First, we discuss results for the roughness of the
growing ledge
on an infinitely long terrace ($L_y = \infty$)
with a large value of $L_x=10^4$. In this case,
after a short initial transient the undulations of the ledge grow and
finger-like structures emerge, separated by deep cracks.
The cracks deepen and the fingers themselves
coarsen at the expense of other fingers.
In Fig. 4, we show a sequence of typical successive configurations for
different values of $\ell_d$. We find that 
the width $w(t)$ of the interface associated with the
ledge follows power law behavior

\begin{equation}
\label{wt}
 w(t)\equiv \langle \overline{[h(x,t)- \bar{h}(t)]^2} \rangle^{1/2}
= A t^{\beta_1},
\end{equation}

where the brackets and the overbar denote an average over the configurations
and over each finite system, respectively. The height variable
$h(x,t)$ is the column height of the ledge as measured from $y=0$.
Numerically, we find that 
the width $w(t)$ grows linearly with $\beta_1 = 1.0\pm 0.01$ and
its slope $A(\ell_d)$ depends on the drift $\ell_d$ (Fig. 5). 
Linear growth can be understood qualitatively, since particles arriving
at the vertical section of the ledge do not contribute significantly to
the ledge roughness. The increase in the roughness
is mainly due to
particles that stick on top of the columns, and thus the width
grows proportional to the total particle number, {\it i.e.} time.
The change in the growth rate is also easy to explain qualitatively.
With small drifts, only the top of the finger grows and very few particles
reach the bottom of the cracks. With larger drifts, the probability
of reaching the bottom increases,
and thus $w$ increases more slowly (see also Fig. 4).

The value of $\beta_1=1$ is 
consistent with the theory of Elkinani and
Villain \cite{Elk94} for a simple 1D Zeno model of MBE
growth with step edge barriers. 
Instead of ledges, they consider
deposition of adatoms on a stepped surface with diffusion.
They show that with strong step edge barriers, deep cracks are formed
on the surface whose depth grows linearly in time. In this
case, the deposition noise is not relevant and this result
can be obtained from a deterministic model. 

To study the effect of a finite terrace length $L_y<\infty$,
we have simulated the model with $\ell_d=1/4$,  $L_x=10^4$, and
$L_y=$ 50, 70, 100, 140, 200, and 500.
Due to the fact that in such finite systems
the relative proportion of the flux deposited in between the
fingers increases with time, the width $w(t)$ eventually saturates to
an $L_y$ dependent value, but does {\it not}
saturate as a function of $L_x$.
We find that the width satisfies 
the scaling {\it ansatz} of Family-Vicsek \cite{Bar95,Fam85}:

\begin{equation}
\label{wlyt}
w(L_y,t)= t^{\chi_1/z_1} f_1(L_y/t^{1/z_1}),
\end{equation}

where the scaling function $f_1(x)$ behaves as

\begin{equation}
\label{f1}
 f_1(x) \sim \cases{\hbox{\it const.} &$x\gg 1$;
                \cr x^{\chi_1} &$x\ll 1$. \cr}
\end{equation}

The exponent $\chi_1$ characterizes the surface morphology
in the saturated regime  $w(L_y) \sim L_y^{\chi_1}$,
and the crossover time $t_{\rm sat} \sim L_y^{z_1}$
determines where the saturation takes over.
The growth exponent for $t \ll t_{sat}$ is $\beta_1= \chi_1/z_1$.
We find that setting $\beta_1=1$, 
$z_1=1.00 \pm 0.03$ collapses
our data best to a single scaling function shown in Fig. 6.
We have also obtained the exponent $\chi_1$ 
by fitting to the saturated width $w(L_y)$ 
and find that $\chi_1 = 0.96 \pm 0.02$.

\subsection{Finger Coarsening}

In our model diffusion along the
ledge is limited by infinitely strong barriers, since the
particles can never go around corners. This is basically the
same effect as step barriers along the surface of
the 1D Zeno model.
However, since in our model there is a real diffusion field
surrounding the fingers on the terrace, 
additional {\it finger coarsening} \cite{Pol96} takes place as
is evident in the configurations of Fig. 4. 
For a finite system with $L_x < \infty$, this eventually leads
to a configuration where there is only one finger present.
To investigate the temporal scaling of the
thickness of the fingers, 
we have studied how the first zero of the Green's function at $r=r_{0}(t)$

\begin{equation}
\label{Grt}
G(r,t) =\langle  \frac{1}{N} \sum_{x} 
h(x+r,t) h(x,t) - \bar{h}(t)^2
\rangle
\end{equation}

\noindent
behaves as a function of time. The behavior of $r_0(t)$ should
indicate the existence of
a characteristic, time-dependent
correlation length in the direction perpendicular
to the direction of growth.
In Fig. 7 we show $r_0(t)$ for several
values of $\ell_d$ when $L_x=10^4$ and $L_y=\infty$.
To a good degree of accuracy, we find
that $r_0(t) \sim t^{\beta_r}$, with the value $\beta_r=0.32 \pm 0.01$
for drifts varying from $1/8$ to $32$. There is, however, a 
long crossover regime at the beginning of the growth that
depends on the drift,
being longer for larger drift values.

It is also interesting to study the scaling of the Green's function.
Asymptotically, we expect $G(r,t)$ to scale as \cite{Bra94}

\begin{equation}
\label{Grtscaling}
G(r,t)=t^{-2\beta_1}g_{\ell_d}(t^{-\beta_r}r),
\end{equation}

where $g_{\ell_d}(x)$ is a new scaling function associated with the
coarsening process.
In Fig. 8 we show scaling of the data for $G(r,t)$,
with a very good
data collapse obtained with $\beta_1=1$ and $\beta_r=1/3$
\cite{scaling}.
It is interesting to note that the finger coarsening
in the present model follows the same power law of 1/3 as
Model B,
which describes domain coarsening due to long range diffusion
\cite{Bra94,Rog88}.
However, although qualitatively similar,
the present scaling function depicted in Fig. 8 is quantitatively different
from that of Model B \cite{Rog88}.
The exponent 1/3 also appears in models of noise-driven coarsening
of mounds in 1D surface growth where slope selection occurs because 
of step edge barriers \cite{Kru93a,Kru97}.

The finite-size scaling of $r_0$ is different from that of the width
$w$, since for a system with a finite terrace width $L_x < \infty$
but with  $L_y=\infty$, the late-time
configuration consist of one finger only, whose vertical
roughness $w$ keeps on growing linearly but whose
$r_0$ saturates. This introduces a new exponent $z_2$ that
controls the saturation of $r_0$ in the $x$ direction.
On the other hand, for $L_x=\infty$ and
$L_y < \infty$, {\it both} $w$ {\it and} $r_0$ saturate, and
their saturation must be characterized by the same exponent
$z_1$ in Eq. (\ref{wlyt}). Thus, for the general case of
both $L_x,L_y < \infty$, we expect the following scaling form
to hold:

\begin{equation}
\label{r0lxlyt}
 r_0(L_x,L_y,t)= t^{\beta_r} f_r(
{L_x\over{t^{1/z_2}}},
{L_y\over{t^{1/z_1}}} ).
\end{equation}

We will not study the whole scaling function $f_r(x,y)$
here but consider
the effects of a finite $L_x$ and $L_y$ separately \cite{needlelimit}.
For $L_y=\infty$, we can again write down the Family-Viscek form
as

\begin{equation}
\label{r0lxt}
 r_0(L_x,t) = t^{\chi_2/z_2} f_2(L_x/t^{1/z_2}),
\end{equation}

where now $\beta_r \equiv \chi_2/z_2$, and the scaling function 
$f_2$ has the same limits as
$f_1$, but now with a new roughness exponent $\chi_2$. We have
simulated the model with 
$L_x=$ 20, 50, 100, and 200 using the drift $\ell_d =1$.
Because of the single finger final configuration, there are
large fluctuations in the data and thus we have determined
the saturation exponent $\chi_2$ 
by estimating the saturated width $r_{0}(L_x)$
directly for various values of 
$L_x$. From the data, our best estimate is $\chi_2 = 1.02 \pm 0.01$,
{\it i.e.} the width of the final finger grows as the horizontal
system size. Together with $\beta_r=0.33$ this implies that
$z_2=3.0$.

In the case of a finite $L_y$, we expect that the scaling 
form satisfies

\begin{equation}
\label{r0lyt}
 r_0(L_y,t) = t^{\chi_3/z_1} f_3(L_y/t^{1/z_1}),
\end{equation}

where $\beta_r$ must now satisfy the relation
$\beta_r=\chi_3/z_1$, with $\chi_3$ being another new roughness
exponent. The limits of 
$f_3$ and $f_1$ are again of the same form. By using system sizes
$L_y=$ 50, 70, 100, 140, 200, and 500 with the drift $\ell_d =1/4$, our
data collapses to the scaling form shown in Fig. 9 
with $\chi_3=0.33 \pm 0.01$
and $z_1=1.00 \pm 0.03$.
Moreover, we have obtained another estimate 
of the new saturation exponent $\chi_3$ 
by estimating the saturated width $r_{0}(L_y)$ 
and indeed verify that $\chi_3=0.34\pm 0.02$.

\section{Summary and Conclusions}

In summary, we have in this work introduced and
examined a very simple model for the
growth of an isolated step with infinitely strong step
edge barriers. The destabilizing effect of the one-sided
biased diffusion field coupled with strongly anisotropic
adatom dynamics makes the ledge morphologically unstable,
with finger-like structures developing separated by deep cracks.
After an initial early-time transient 
the fingers coarsen as $t^{0.33}$ and the width of the 
ledge grows linearly. 
For an infinitely wide and long terrace, the fingers eventually
become needle-like. We have also studied the finite-size scaling
of both the coarsening and the width of the ledge in detail,
and determined the corresponding scaling exponents.

Recently, Pierre-Louis {\it et al.} \cite{Pie97} have considered
in detail a more realistic model of step train
growth in the case of weak desorption,
and one-side attachment. As in the present case, they find that
the step morphology is linearly unstable, but now the individual
step widths grow $\propto t^{1/2}$, with the steps ``locked in''
together. In this regime, there is no step coarsening, either.
Thus, we expect our model to be relevant only for the case where
the steps are well isolated, and detachment from step edges 
can be neglected.

Experimentally, growth of steps on 
slightly miscut Si(001) surfaces has been studied,
with the claimed result that the step roughening is consistent
with the KPZ prediction \cite{Wu93}. However, at least
superficially 
the steps depicted in Ref. \cite{Wu93} appear to develop
finger-like structures separated by deep grooves
characteristic of the unstable regime studied here
and in Ref. \cite{Pie97}.
It would be interesting to carry out more systematic studies
of roughening of widely spaced steps on semicondutor surfaces to 
characterize the nature of the instability.

Acknowledgements: T. A-N. wishes to thank J. Krug and
M. Rost for useful discussions. 
This work has in part been supported by the Academy of Finland,
and by NSF grant No. DMR - 9222812.

$^*$Corresponding author. E-mail address: {\tt alanissi@csc.fi}.

\appendix
\section*{}

In this Appendix, we calculate explicitly 
the average arrival time $t_{\rm arr}$ of a walker 
to demonstrate the need to use the present algorithm.
We will also discuss the convergence of the probability
distribution for finite terraces with $L_y < \infty$.
To begin with, the distribution for the number of steps or
the distribution of
the first passage time $P_{Ly}(t)$, 
can be calculated similarly to the
spatial distribution of Eq. (\ref{Plyx}) by the recursion

\begin{eqnarray}
\label{Plyt}
P_{Ly}(t) & = & {1\over 2a}
\sum_{s=1}^{\infty}
\delta_{t-s,1}
[b \delta_{s,0}+2P_{Ly}(s)+
\nonumber \\ 
& & \sum_{u=1}^{\infty}P_{Ly}(s-u)P_{Ly-1}(u)].
\end{eqnarray}

Using the temporal Fourier transform

\begin{equation}
\widetilde P_{Ly}(\omega)=\sum_{t=-\infty}^{\infty} e^{-i \omega t}P_{Ly}(t)
\end{equation}

we obtain the characteristic function as

\begin{equation}
\label{Plyomega}
\widetilde P_{Ly}(\omega)={b \over 2ae^{i\omega}- 2 - \widetilde
P_{Ly-1}(\omega)},
\end{equation}

which can be solved with the initial value
$\widetilde P_0(\omega)=1$. Again, 
$\widetilde P_{Ly}(\omega)=\widetilde P_{Ly-1}(\omega)$ when
$L_y\rightarrow\infty$ and $\widetilde P_{\infty}(\omega)$ can be obtained.
The average arrival time is proportional to the first derivative of
the characteristic function at $\omega=0$ by

\begin{eqnarray}
t_{\rm arr} & = & \sum_{t=1}^{\infty}tP(t)
= i{{d \widetilde P(\omega)}\over{d\omega}}\vert_{\omega=0}
\nonumber \\ 
& = & -{\widetilde P^2_{Ly}(0) \over b}
[2ia- {{d \widetilde P_{Ly-1}(\omega)}\over{d\omega}}\vert_{\omega=0}]
\nonumber \\ 
& =  & {{d \widetilde P_{Ly}(\omega)}\over{d\omega}}\vert_{\omega=0}=
-2ia \sum_{n=1}^{Ly} ({ {1}\over{b} })^n
\nonumber \\ 
& = & (1+{1\over{\ell_d}})[1-{1\over{(1+4\ell_d)^{Ly}}}]
\end{eqnarray}

when $\ell_d > 0$. In the infinite terrace limit ($L_y
\rightarrow \infty$), $t_{\rm arr} = 1 + 1/\ell_d$,
while for $\ell_d=0$ it is easy to show that
$t_{\rm arr} = 4L_y$.
Thus, the return time quickly becomes prohibitively large
for large systems and small values of the drift, making
brute-force Monte Carlo simulations difficult.
On the other hand, for drifts larger than unity,
no significant reduction in computer time can be obtained
with the new algorithm.

Finally, to estimate the convergence of the probability
distribution $P_{Ly}(x)$ towards its asymptotic limit
as a function of the terrace length $L_y$ for any
$L_x$, we can define the deviation $d$ by

\begin{eqnarray}
& & d^2(L_x,L_y,\ell_d)
\nonumber \\ 
& & ={1\over{L_x}}\sum_{n=0}^{L_x-1}
[ \widetilde P_{Ly}({{n}\over{2 \pi L_x}})-
  \widetilde P_{\infty}({{n}\over{2 \pi L_x}}) ]^2,
\end{eqnarray}

assuming a periodic system in the $x$ direction. In Fig. 10,
we show the deviation for various values of the drift as
a function of $L_y$.

%

\vskip1.0cm

\begin{figure}
\caption{
  A schematic view on adatom dynamics on
  a vicinal surface. The local adatom
  concentration is denoted by $n$ and
  the other symbols are explained in text.
}
\label{}
\end{figure}

\medskip
\begin{figure}
\caption{
  Adatom dynamics in the growth model. For $L_y < \infty$, deposition
  occurs uniformly randomly at all available (unoccupied) sites, while
  for $L_y = \infty$, the particles are released from the line
  at $y=H+1$. Particles then
  diffuse on the lower terrace, and
  drift in the $-y$ direction. They become incorporated into
  the step when they either land on a top section of
  the step (along the $x$ axis),
  or slide down along the $-y$ direction
  to the nearest kink site, as shown schematically in the figure.
  If a particle attempts to cross the line at $(x_0,H+1)$ in the
  $y$ direction, it is immediately returned from ($x_0,H+2$) to the line
  with new coordinates ($x_0+x,H+1$) chosen from
  the spatial distribution $P_{Ly}(x)$.
  The boundary above $y=H+L_y+1$ (not shown) is
  completely reflecting.
}
\label{}
\end{figure}

\medskip
\begin{figure}
\caption{
  Probability distribution $P_{\infty}(x)$ for the return position
  with $L_x=10^4$ and $L_y=\infty$. The drift parameter
  $\ell_d=$ 1, 1/2, 1/4, 1/8, 1/16, and 0 (from top to bottom at $x=0$).
  In the limit $\ell_d \rightarrow \infty$ the distribution approaches
  a delta function.
}
\label{}
\end{figure}

\medskip
\begin{figure}
\caption{
  Ten consecutive step profiles from the growth model
  with $L_x=10^4$ and $L_y=\infty$
  at $t=1000, 2000,...,10^4$ with the drift
  (a) $\ell_d=1/4$,
  (b) $\ell_d=1$, and
  (c) $\ell_d=4$. Only part of the system is shown.
}
\label{}
\end{figure}

\medskip
\begin{figure}
\caption{
  The step width $w(t)$
  for $\ell_d=$ 1/8, 1/4, 1/2, 1, 2, 4, 8, 16, and 32 (from top to bottom)
  with $L_x=10^4$ and $L_y=\infty$.
  The slope $A$ as a function of the drift $\ell_d$
  is shown in the inset.
 }
\label{}
\end{figure}

\medskip
\begin{figure}
\caption{
  Scaling function $f_1$ of Eq. (10) for the step width $w(L_y,t)$
  with $L_y=$ 50, 70, 100, 140, 200, and 500. Good scaling is
  obtained with $\beta_1=1$ and $z_1=1$.
  The drift $\ell_d=1/4$ and the lateral lattice size $L_x=10^4$.
 }
\label{}
\end{figure}
\medskip
\begin{figure}
\caption{
  The finger width $r_0(t)$
  for $\ell_d=$ 1/8, 1/4, 1/2, 1, 2, 4, 8, 16, and 32 (from top to bottom)
  with $L_x=10^4$ and $L_y=\infty$.
  The dashed line indicates a slope of $1/3$.
 }
\label{}
\end{figure}

\medskip
\begin{figure}
\caption{
 Scaling function $g_{\ell_d}$ 
of Eq. (13) for $G(r,t)$ with $L_x=10^4$ and $\ell_d=1$
 at ten different times $t=1000, 2000,...,10^4$, with
 $\beta_1=1$ and $\beta_r=1/3$.
 }
\label{}
\end{figure}

\medskip
\begin{figure}
\caption{
  Scaling function $f_3$ of Eq. (16) for the finger width $r_0$
  with the terrace length $L_y=$ 50, 70, 100, 140, 200, and 500.
  The drift $\ell_d=1/4$ and the lateral lattice size $L_x=10^4$,
  and $\beta_r=1/3$ and $z_1=1$.
 }
\label{}
\end{figure}

\medskip
\begin{figure}
\caption{
  Root mean square deviation $d$ of the propagator $P_{Ly}$ from
  $P_{\infty}$ with $L_x=10^4$ for
  $\ell_d=$ 0, 1/16, 1/8, 1/4, and 4 (from top to bottom),
  shown as a function of the distance $L_y$ to the reflecting boundary.
 }
\label{}
\end{figure}


\begin{references}
\bibitem{drift}
In the case of one-particle diffusion, the drift term is not
strictly speaking present since the step does not move. 
In our model $\ell_d$ may be considered as a finite-density
correction to this limit.
\bibitem{irrelevant}
On a Si(001) terrace diffusion
along the dimer rows is much faster than across them.
For $\ell_d > 0$, however, the
step morphology is controlled by the drift term, and thus the 
exact value of the ratio $D_{xx}/D_{yy}$ is not important.
\bibitem{scaling}
As defined in Eq. (13), the scaling function $g_{\ell_d}(x)$ depends
on the drift $\ell_d$. We can also write the scaling form as
$G(r,t) = w^2(\ell_d,t) g(r/r_0(\ell_d,t))$, in which case
the scaling plots for the new function
$g(x)$ collapse within the accuracy of
our data for the values of the drift studied, namely
$\ell_d=1/8$, 1/4, 1/2, and 1.
\bibitem{needlelimit}
In the long-time limit,
for an infinite system ($L_x=L_y=\infty$) the
aspect ratio $r_0/w$ approaches zero, and the fingers
become needle-like even for finite values of $\ell_d$. 
In the needle model, the
penetration probability of the
diffusing particles between needles decays exponentially as
$\exp(-\pi r /\xi)$ \cite{Kru93b}, where the distance $r$ is
measured from the top of the needles and $\xi$ is the distance
between two needles.
In our model with a finite drift parameter $\ell_d$, 
the corresponding result can be shown to be 
$\exp{ \{ - \frac{1}{2D_{yy}} [ \sqrt{ \frac{4D_{xx}D_{yy}\pi^2}{\xi^2}
+ v_d^2(\ell_d)} +  v_d(\ell_d) ] r \} }$, where $v_d(\ell)$,
$D_{xx}$, and $D_{yy}$ are defined in the text. For zero drift
(where $v_d=0$, and $D_{yy}=D_{xx}$),
this correctly reduces to the previous result.
\end{references}
\end{document}